# Strain-tunable Dirac semimetal phase transition and emergent superconductivity in a borophane

Chengyong Zhong [1 ✉], Xuelian Li[1] & Peng Yu [1 ✉]

A two-dimensional (2D) Dirac semimetal with concomitant superconductivity has been long sought but rarely reported. It is believed that light-element materials have the potential to realize this goal owing to their intrinsic lightweight and metallicity. Here, based on the recently synthesized $\beta_{12}$ hydrogenated borophene, we investigate its counterpart named $\beta_{12}$-$B_5H_3$. Our first-principles calculations suggest it has good stability. $\beta_{12}$-$B_5H_3$ is a scarce Dirac semimetal demonstrating a strain-tunable phase transition from three Dirac cones to a single Dirac cone. Additionally, $\beta_{12}$-$B_5H_3$ is also a superior phonon-mediated superconductor with a superconducting critical temperature of 32.4 K and can be further boosted to 42 K under external strain. The concurrence of Dirac fermions and superconductivity, supplemented with dual tunabilities, reveals $\beta_{12}$-$B_5H_3$ is an attractive platform to study either quantum phase transition in 2D Dirac semimetal or the superconductivity or the exotic physics brought about by their interplay.

[1] College of Physics and Electronic Engineering, Chongqing Normal University, 401331 Chongqing, China. ✉email: zhongcy@cqnu.edu.cn; pengyu@cqnu.edu.cn

In the past years, 2D materials have attracted considerable attention not only because they provide the possibility to easily control their properties by external methods[1], but also because of the plenty of salient physical properties brought about by their reduced dimensionality and electron confinement[2,3]. Among various properties, Dirac semimetallicity and superconductivity are two active topics[4–6]. In general, one material cannot possess Dirac properties around its Fermi level and superconductivity simultaneously, since the Fermi surface (FS) of a Dirac semimetal usually consists of discrete points or nodal lines, which conflicts with the requirement of a high carrier density near the Fermi level to give rise to a high critical temperature ($T_c$). For instance, the intrinsic graphene (the first proposed 2D Dirac semimetal) is not superconducting and becomes a superconductor that needs the aid of doping and external strain[7]. The coexistence of Dirac properties and superconductivity possibly lures many exotic properties[8,9], therefore, it is highly interesting to search for this kind of 2D materials harboring these two properties.

The band crossing points in Dirac semimetals are typically more robust in light-element (e.g. boron and carbon) materials owing to their intrinsic negligible spin-orbital coupling strengths, as exemplified by dozens of Dirac semimetals have been proposed in boron[10–12] and carbon materials[13–15]. On the other hand, according to the conventional Bardeen-Cooper-Schrieffer (BCS) theory[16], it is believed that lightweight metals have a better chance to induce high $T_c$, because the Debye temperatures within these metals are usually high enough to induce strong phonon-mediated superconducting pairing. Thus, it is natural to prospect a 2D material with both Dirac and superconducting properties has a higher opportunity to be found in light-element materials. Scanning the periodic table, it seems that boron is a desirable candidate that can combine lightweight and metallicity in a 2D material to simultaneously trigger Dirac and superconducting properties. In fact, this assumption has been partially confirmed in the 2D boron materials (i.e. borophenes). For example, the Dirac fermions[17,18] and superconductivity[19,20] have been experimentally or theoretically verified in the synthesized borophene $\beta_{12}$ and $\chi_3$, the backbone of honeycomb borophene in the well-known superconductor MgB$_2$ is also mainly responsible for the appearance of its Dirac nodal line[21] and high $T_c$[22,23].

2D borophenes have wide potential applications[24–29]. However, the bare surface makes the borophenes susceptible and easy to be oxidized at ambient conditions[30]. Chemical functionalization, such as hydrogenation, usually serves as an effective knob to remedy this problem and while be regarded as a desirable approach to modulate their different properties[31]. It is worthwhile to point out that hydrogenated borophenes (also known as borophanes) are also suitable for spawning Dirac fermions and superconductors since hydrogen and boron are both light elements. Hydrogenated borophenes have been theoretically found to be highly stable and possess ideal Dirac cones[32–35] or Dirac nodal loops[36]. A honeycomb borophene hydride (named as h-B$_2$H$_2$ here) was successfully achieved by exfoliation and complete ion exchange between protons and magnesium cations in MgB$_2$[37], and a hydrogenated borophene based on $\beta_{12}$ (named as $\beta_{12}$-B$_5$H$_2$ here) was also experimentally realized on the substrate[38]. The h-B$_2$H$_2$ is a normal metal with poor superconductivity[39] and the existence of free-standing $\beta_{12}$-B$_5$H$_2$ is still questionable (we will discuss it later). It is natural to wonder whether there exists a stable hydrogenated borophene with superb superconductivity, or clean Dirac elements or even both.

The answer is affirmative, in this work, based on synthesized borophene $\beta_{12}$ [see Fig. 1a, b] and hydrogenated $\beta_{12}$ borophene $\beta_{12}$-B$_5$H$_2$ [see Fig. 1c, d], we propose a hydrogenated $\beta_{12}$ borophene named as $\beta_{12}$-B$_5$H$_3$ [see Fig. 1e, f] can fulfill these expectations. First, our first-principles calculations of mechanical properties and phonon spectrum suggest $\beta_{12}$-B$_5$H$_3$ exhibits good stability, even better than its brother $\beta_{12}$-B$_5$H$_2$. Second, $\beta_{12}$-B$_5$H$_3$ is a Dirac semimetal with three clean Dirac cones (two type-I Dirac cones and one type-II Dirac cone) near the Fermi level at its equilibrium and can be tailored into a Dirac semimetal with a single type-I Dirac cone under external strain beyond 3.8% along the **a** direction. To our knowledge, such a feature has not been found before. In the end, based on the anisotropic Migdal-Eliashberg (ME) equations, we find that the $T_c$ of $\beta_{12}$-B$_5$H$_3$ is as high as 32.4 K and can be boosted to 42 K at 5.8% external strain along the **b** direction. Therefore, $\beta_{12}$-B$_5$H$_3$ is a Dirac semimetal with triple Dirac cones and while with superconducting features, which make it a platform to study the exotic physics brought about by either Dirac points or superconductivity or both of them.

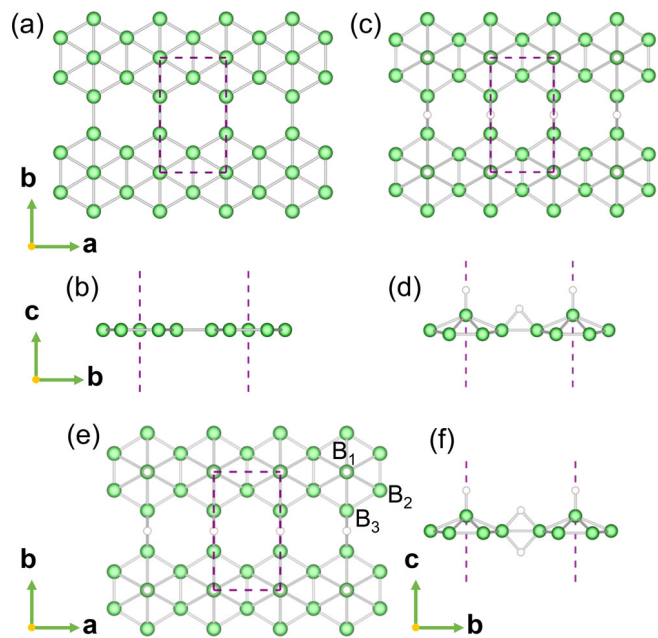

**Fig. 1 Crystal structure.** (**a**) Top view and (**b**) side view of $\beta_{12}$. **c** Top view and (**d**) side view of $\beta_{12}$-B$_5$H$_2$. **e** Top view and (**f**) side view of $\beta_{12}$-B$_5$H$_3$. Green and pink balls represent boron and hydrogen atoms, respectively. The dashed purple rectangles indicate their unit cells.

## Results

**Atomic structure and mechanical properties.** Due to the nature of electron-deficient in boron, the 2D borophenes demonstrate various polymorphs with different hole densities referring to the triangular lattice[40]. The hole density of $\beta_{12}$ [see Fig. 1a] is 1/6 making the whole electrons a little surplus in terms of the electron counting rule[41], which partially explains why it would be quickly oxidized at ambient conditions. In a chemical sense, the excess electrons can be neutralized by chemical passivation, such as hydrogenation, and thus improve stability. The recent experiment has confirmed that the oxidation rate can be reduced by more than two orders of magnitude after hydrogenating $\beta_{12}$ borophene to a well-ordered borophane on the substrate[i.e. $\beta_{12}$-B$_5$H$_2$, see Fig. 1c, d][38]. According to our first-principles calculations, the free-standing $\beta_{12}$-B$_5$H$_2$ is not stable since there are sizable soft modes in its phonon spectrum [see Supplementary Fig. 1a–d and more discussion about their stabilities can be found in Supplementary Note 1]. However, this phonon instability can be eliminated by adding an enantiomer of the bridge hydrogen in $\beta_{12}$-B$_5$H$_2$ to form a hydrogenated borophene called as $\beta_{12}$-B$_5$H$_3$ here [see Fig. 1e, f], which its phonon spectrum shows no any soft modes throughout the entire Brillouin Zone (BZ) [see Supplementary Fig. 1e, f]. The optimized lattice constants of $\beta_{12}$-B$_5$H$_3$

Table 1 The structural and mechanical properties.

| System | Space Group | a | b | $C_{11}$ | $C_{22}$ | $C_{12}$ | $C_{44}$ | $Y_a$ | $Y_b$ | $\nu_a$ | $\nu_b$ |
|---|---|---|---|---|---|---|---|---|---|---|---|
| $\beta_{12}$[42] | Pmmm | 2.93 | 5.07 | 214.3 | 188.1 | 36.0 | 63.5 | 207.5 | 182.0 | 0.17 | 0.19 |
| $\beta_{12}$ [This work] | Pmmm | 2.93 | 5.06 | 218.6 | 187.1 | 36.8 | 62.7 | 211.4 | 180.9 | 0.17 | 0.19 |
| $\beta_{12}$-$B_5H_2$ | Pmm2 | 2.84 | 5.12 | 109.8 | 176.5 | 41.2 | 62.6 | 100.2 | 161.0 | 0.38 | 0.23 |
| $\beta_{12}$-$B_5H_3$ | Pmm2 | 2.90 | 5.16 | 153.5 | 181.9 | 25.4 | 70.3 | 149.9 | 177.7 | 0.17 | 0.14 |
| h-$B_2H_2$ | Cmmm | 3.02 | 5.29 | 83.7 | 140.6 | 37.4 | 55.6 | 73.9 | 123.9 | 0.45 | 0.27 |

The space groups, lattice constants (Å), elastic constants (N/m), Young's moduli (N/m) and Poisson ratios for $\beta_{12}$, $\beta_{12}$-$B_5H_2$, $\beta_{12}$-$B_5H_3$ and h-$B_2H_2$, respectively.

are $a = 2.90$ Å and $b = 5.16$ Å, varying not too much in comparison with those of its brother $\beta_{12}$-$B_5H_2$ and mother $\beta_{12}$ (see Table 1). Three different kinds of boron atoms are marked as $B_1$, $B_2$ and $B_3$, respectively [see Fig. 1e]. In experiments, $\beta_{12}$-$B_5H_3$ can be prepared by hydrogenating $\beta_{12}$ borophene or $\beta_{12}$-$B_5H_2$ via in situ and three-step thermal-decomposition process as conducted in the previous work[31] once the free-standing $\beta_{12}$ borophene or $\beta_{12}$-$B_5H_2$ is produced.

It has been reported that $\beta_{12}$ has good mechanical properties[42], we are also curious about its performance in $\beta_{12}$-$B_5H_3$. The calculated independent elastic constants of $\beta_{12}$-$B_5H_3$ are $C_{11} = 153.5$ (N/m), $C_{22} = 181.9$ (N/m), $C_{12} = 25.4$ (N/m) and $C_{44} = 70.3$ (N/m), which obviously satisfy the Born-Huang mechanical stable criteria: $C_{11}C_{22} - C_{12}^2 > 0$ and $C_{44} > 0$. The small difference of Young's moduli and Poisson ratios of $\beta_{12}$-$B_5H_3$ along **a** ($Y_a = 149.9$ N/m, $\nu_a = 0.17$) and **b** ($Y_b = 177.7$ N/m, $\nu_b = 0.14$) directions indicate its mechanical anisotropy is not prominent [see Table 1, more details can be found in Supplementary Note 2 and Supplementary Fig. 3]. Although the elastic constants of $\beta_{12}$-$B_5H_3$ are little smaller than those of $\beta_{12}$ but better than those of synthesized h-$B_2H_2$ (see Table 1). We have also calculated the strain-stress curves along the **a** and **b** directions [see Supplementary Fig. 4], which shows that $\beta_{12}$-$B_5H_3$ is within linear elastic regime up to at least 6% along the two directions, and its elastic strain limits are high(tensile strain limits are even beyond 20% along the two directions). The excellent mechanical properties of $\beta_{12}$-$B_5H_3$ would facilitate the application of external strain for tuning its various properties.

**Strain-tunable Dirac semimetal phase transition**. The most fascinating feature of $\beta_{12}$-$B_5H_3$ lies in its electronic properties. The electronic band structure for $\beta_{12}$-$B_5H_3$ is shown in Fig. 2a. Notably, it can be observed that two bands cross each other three times (two times along Γ–X and one time along Γ–Y) to form three Dirac points near the Fermi level as indicated with green, red and blue circles in Fig. 2a. Their corresponding positions in BZ are sketched in Fig. 2g. Depending on the type of band dispersion, a Dirac point can be classified as type-I and type-II[43]: a type-I Dirac point is formed by the crossing between an electron-like band and a hole-like band; a type-II Dirac point is formed by the crossing between two electron-like bands or hole-like bands. Here, after examining the dispersions around these three Dirac points, we find that two of them are type-I Dirac points [see Fig. 2c, e] and the other one is a type-II Dirac point [see Fig. 2d], which can also be inferred from the three-dimensional (3D) band structure [see Fig. 2f]. Another highlight worth being pointed out is that there are only two Dirac bands in a broad energy range around the Fermi level, which should be beneficial for experimentally detecting these Dirac fermions. We further check the band structure by the more accurate hybrid functional method [see Supplementary Fig. 5], which confirms that the key band features are still maintained. Therefore, the following discussion would be based on the PBE results.

The orbital-resolved band structure denotes the two Dirac bands are mainly contributed by $p_x$ and $p_z$ orbitals of boron, which are further consolidated by the projected density of states [PDOS, see Fig. 2b.] The two Dirac bands form two elliptical electron Fermi sheets along the **a** direction and two triangular-like hole Fermi sheets along the **b** direction [see Fig. 2h]. These Fermi sheets also mainly originated from $p_x$ and $p_z$ orbitals of boron [see Fig. 2h] and demonstrate strong anisotropy.

The space group of $\beta_{12}$-$B_5H_3$ is Pmm2 (NO. 25), the little group along Γ–X/Y is $C_s$ containing an identical symmetry (E) and a mirror symmetry ($\sigma_h$). According to our calculations, the two Dirac bands have opposite mirror eigenvalues and thus belong to different irreducible representations [see Fig. 2c–e], which means the two Dirac bands cannot hybridize and the crossing between them cannot be gapped.

In the original $\beta_{12}$ borophene, its FS consists of $p_x$, $p_y$ and $p_z$ orbitals of boron[20] and external strains cannot change its strong metallicity[44,45]. In contrast, the Fermi sheets of $\beta_{12}$-$B_5H_3$ are contributed by $p_x$ and $p_z$ orbitals of boron and three Dirac points are formed around the Fermi level. We are curious about how the electronic properties of $\beta_{12}$-$B_5H_3$ are affected by the external strains. The appearance of the three Dirac points is attributed to the two times band inversions between the valence band and conduction band. We denote the magnitude of the band inversion that occurred at Γ is Δ [see Fig. 3a]. In Fig. 3d, we plot the Δ as a function of tensile strain along the **a** direction. One observes that there exists a critical strain of 3.8%, during which the band order is switched [see Fig. 3b]. Furthermore, beyond the critical into a single Dirac point phase [see Fig. 3c]. Since the mirror symmetry is preserved during the tensile strain along the **a** direction, the single Dirac point in Fig. 3c is still protected by the mirror symmetry. Thus, a quantum phase transition from Dirac semimetal with triple Dirac points into a Dirac semimetal with a single Dirac point is realized in $\beta_{12}$-$B_5H_3$ by applying tensile strain along the **a** direction.

It is widely accepted that the type-I Dirac fermion in graphene is responsible for its various exotic properties, such as high carrier mobility, Klein tunneling and some quantum behaviors[46]. Type-II Dirac fermion however can give rise to many properties, such as a direction-dependent chiral anomaly[47], an antichiral effect of the chiral Landau level[48] and quantum oscillations due to momentum-space Klein tunneling[49]. The concurrence of type-I and type-II Dirac fermions in the $\beta_{12}$-$B_5H_3$ provides an opportunity to observe how these properties coexist or whether they will coherent with each other. On the other side, the strain-tunable Dirac phase transition from triple Dirac fermions to a single Dirac fermion offers a platform to study how these exotic properties evolve in the course of this phase transition.

**Isotropic and anisotropic superconducting properties**. Generally speaking, a Dirac semimetal with a single Dirac point sitting rightly at the Fermi level such as in graphene cannot possess superconductivity. As for the six Dirac points in the whole BZ of $\beta_{12}$-$B_5H_3$, they deviate slightly from the Fermi level would not influence their detection in experiments, but expands the six Fermi points to four Fermi loops [see Fig. 2g, h] and thus is an advantage for spawning superconductivity. Moreover, it can be observed that the $p_z$-induced Dirac band of the type-II Dirac point is rather dispersionless and forms a van Hove singularity (vHS), leading to a quite large peak in the PDOS near the Fermi level [see Fig. 2b], which may trigger some exotic physics, such as superconductivity. On the other side, we have justified above that lightweight metal, such as the $\beta_{12}$-B$_5$H$_3$ composed of boron and hydrogen, has a great potential to spawn good superconductivity. Therefore, it is highly interesting to investigate the superconducting properties in the $\beta_{12}$-B$_5$H$_3$.

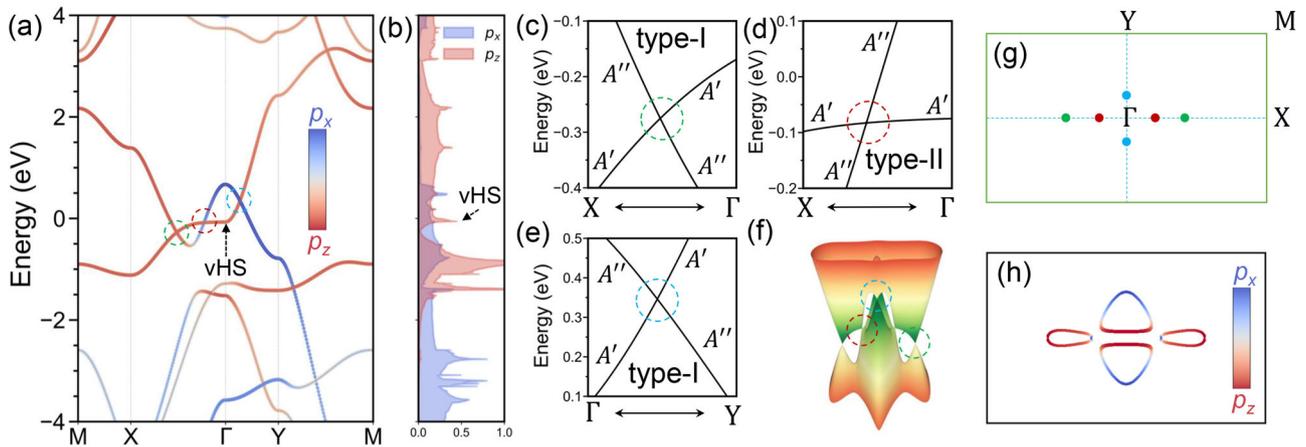

**Fig. 2 The electronic properties of $\beta_{12}$-B$_5$H$_3$. a** The band structure and (**b**) the projected density of states of $\beta_{12}$-B$_5$H$_3$. The band dispersions are coded with orbital characters. The color gradient from blue to red stands for a varying contribution from $p_x$ to $p_z$ orbital of boron. The position of a van Hove singularity (vHS) is indicated with an arrow. **c**–**e** The magnified band structures correspond to the regions circled with different colors in (**a**). **f** The 3D band structure of $\beta_{12}$-B$_5$H$_3$, the colored dashed circles indicate the positions of the Dirac points marked in (**a**). **g** The sketched positions of Dirac points in the Brillouin Zone, the dots with different colors correspond to the different Dirac points marked with the same colors in (**a**). **h** The Fermi surface of $\beta_{12}$-B$_5$H$_3$ upon being projected with $p_x$ and $p_z$ orbitals of boron.

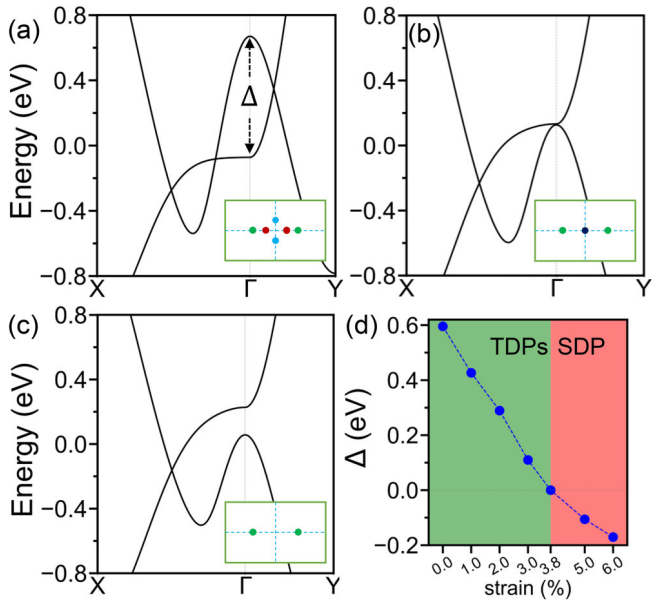

**Fig. 3 The electronic properties of $\beta_{12}$-B$_5$H$_3$ under uniaxial tensile strain along a direction.** Electronic band structures of $\beta_{12}$-B$_5$H$_3$ on the path X–Γ–Y under the applied tensile strain of (**a**) 0%, (**b**) 3.8%, and (**c**) 6%. The band inversion gap at Γ is defined as Δ as marked in (**a**). The insets in (**a**–**c**) indicate the positions of Dirac points in the Brillouin Zone. **d** The Δ as a function of tensile strain along **a** direction. TDPs and SDP are defined as triple Dirac points and single Dirac point, respectively.

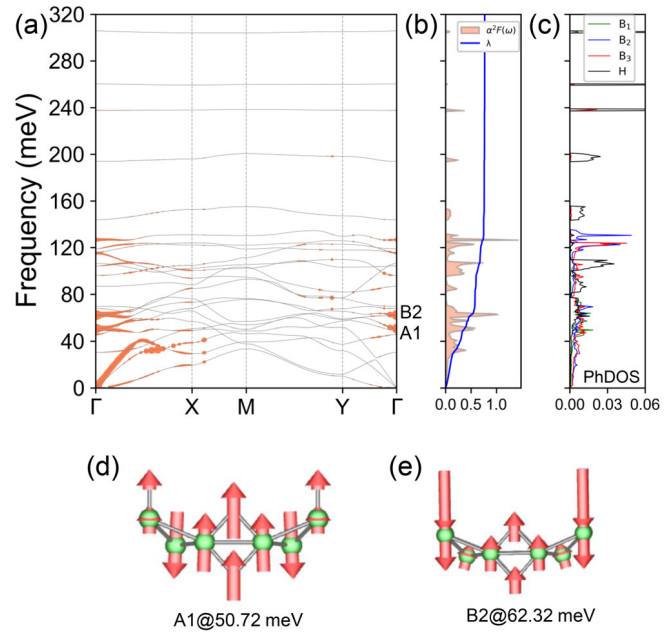

**Fig. 4 The phonon and superconducting properties of $\beta_{12}$-B$_5$H$_3$. a** Phonon dispersions of $\beta_{12}$-B$_5$H$_3$, where the area of the orange dots is proportional to the magnitude of the mode-($\nu$) and momentum-($q$) dependent electron-phonon coupling (EPC) $\lambda_{q\nu}$. **b** The calculated Eliashberg spectral function $\alpha^2 F(\omega)$ and the cumulative EPC strength $\lambda$, where $\omega$ means phonon frequency. **c** The phonon density of states for H, B$_1$, B$_2$ and B$_3$ atoms. Eigenvectors of two modes at (**d**) A1 and (**e**) B2 phonon frequencies as labeled in (**a**). The red arrows and their lengths indicate the directions and relative amplitudes of these two phonon modes, respectively.

We proceed to investigate the vibrational properties and the electron-phonon coupling(EPC) in $\beta_{12}$-B$_5$H$_3$ based on the density-functional perturbation theory[50] as implemented in Quantum ESPRESSO[51]. The dynamic stability of the $\beta_{12}$-B$_5$H$_3$ can be inferred from the phonon band structure, on which the mode-($\nu$) and momentum-($q$) dependent EPC $\lambda_{q\nu}$ are projected [see Fig. 4a]. We find that the phonon modes with low frequencies are key to achieving a high EPC in $\beta_{12}$-B$_5$H$_3$, where about 75% of the total EPC is induced by the phonons with energy lower than 80 meV [see Fig. 4b]. Within this lower energy range, two modes [A1 and B2 modes as indicated in Fig. 4a] have large EPC strengths. A1 phonon mode is an out-of-plane shear mode involving mainly the $B_2$ and $B_3$ atoms, the contribution from the $B_3$ and H atoms is relatively small [see Fig. 4d]. B2 phonon mode is also an out-of-plane shear mode but contributed mostly by the $B_3$ and H atoms instead [see Fig. 4e]. Note that in these two modes, H atoms are involved, which is corroborated by the phonon density of states shown in Fig. 4c, implying that H atoms play an important role in the appearance of superconductivity in $\beta_{12}$-$B_5H_3$. These two modes result in two large peaks in the Eliashberg spectral function $\alpha^2F(\omega)$ [see Fig. 4b].

The Eliashberg spectral function $\alpha^2F(\omega)$ is a central parameter, through which we can obtain the EPC constant $\lambda$ and the logarithmic average frequency $\omega_{log}$ with the following equations:

$$\lambda = \sum_{q\nu} \lambda_{q\nu} = 2\int \frac{\alpha^2F(\omega)}{\omega} d\omega \quad (1)$$

$$\omega_{log} = \exp\left[\frac{2}{\lambda}\int \frac{d\omega}{\omega}\alpha^2F(\omega)\log\omega\right] \quad (2)$$

The calculated $\lambda = 0.77$ and $\omega_{log} = 44.27$ meV for $\beta_{12}$-$B_5H_3$, of which the value of $\lambda$ (0.77) is comparable to that of the well-known phonon-mediated superconductor $MgB_2$ ($\lambda = 0.748$ in the ref. [52]), implying $\beta_{12}$-$B_5H_3$ is also a potential phonon-mediated superconductor with good superconductivity. We can determine the $T_c$ using the McMillian–Allen–Dynes (MAD) formula[53,54] as follows:

$$T_c = \frac{\omega_{log}}{1.20}\exp\left[\frac{-1.04(1+\lambda)}{\lambda - \mu^*(1+0.62\lambda)}\right] \quad (3)$$

where $\mu^*$ is the effective screened Coulomb repulsion constant. For $\beta_{12}$-$B_5H_3$, the calculated $T_c = 21.96$ K (setting $\mu^* = 0.1$), which is a little larger than that of $\beta_{12}$ (18.7 K)[19].

The MAD formula works reasonably well for conventional bulk metals and for weakly anisotropic superconductors such as in bulk lead[55]. However, for layered systems, systems of reduced dimensionality, and those with complex multisheet Fermi surfaces, proper treatment of the anisotropic electron-phonon interaction is required, which has been demonstrated in the well-known $MgB_2$[52]. As mentioned above, the Fermi surface of $\beta_{12}$-$B_5H_3$ is rather anisotropic, therefore, it is believed that the superconducting properties should be more accurate in the formalism of the ME approximation. For $\beta_{12}$-$B_5H_3$, the variation of momentum-dependent EPC parameter $\lambda_k$ and the superconducting gap $\Delta_k$ at 10 K are shown in Fig. 5a, b. Both quantities display similar anisotropy, with their maximum ($\lambda^{max} = 1.12$ meV, $\Delta^{max} = 5.73$ meV) along the Γ–X direction and over double larger than their minimum ($\lambda^{min} = 0.51$ meV, $\Delta^{min} = 2.80$ meV) along the Γ–Y direction, signifying that the anisotropy of the superconducting gap at the FS is strong and the necessity of predicting $T_c$ by the anisotropic ME formula. In addition, compared with the orbital-resolved FS plotted in Fig. 2h, one can see that the relatively strong EPC along the Γ–X direction is mainly contributed by the electronic states on the red section of FS [i.e. the $p_z$-induced bands, see Fig. 2h and Fig. 5a, b]. Figure 5c shows the evolution of the superconducting gap as a function of temperature by solving the ME equations in both the isotropic and the fully anisotropic approximations. The $T_c$ computed within the isotropic approximation is 27.3 K, which is larger than the value (21.96 K) obtained from the MAD formula but smaller than the fully anisotropic result (32.4 K). Within the same anisotropic approximation, the $T_c$(32.4 K) of $\beta_{12}$-$B_5H_3$ is almost identical to the value (33.0 K) of $\beta_{12}$[56].

A Dirac semimetal phase transition has been observed in $\beta_{12}$-$B_5H_3$ under external strain along the the **a** direction, we expect the external strain would have also a great effect on its superconducting properties. We calculate the effects of a tensile strain along the **b** direction up to 5.8% on the EPC $\lambda$, the logarithmic average frequency $\omega_{log}$ and the $T_c$s estimated within three different approximations [i.e. MAD, isotropic ME and anisotropic ME, see Fig. 6a, b]. One observes that $\lambda$ decreases at first and then go arise to 1.42 at 5.8%. The $\omega_{log}$ keeps diminishing and drops down to 16.66 meV at 5.8%. Their combined result is the $T_c$ initially decreases and then slowly increases [see Fig. 6a]. Note that the values of $T_c$ overall becomes smaller, although not too much, after subjecting from tensile strain along the **b** direction in the MAD and isotropic ME approximations, but the $T_c$ calculated with the more accurate anisotropic ME method is boosted to 42 K at 5.8% tensile strain, which is a record in the hydrogenated borophenes. It is revealed that the vibrational modes of H atoms are pushed towards a high-frequency range during the tensile strain along the **b** direction increases and their contributions on the strain-mediated $T_c$ is weakening (more detailed discussion can be found in Supplementary Note 3). Interestingly, we notice that the U-shape of $T_c$ versus tensile strain relationship also occurred in the original $\beta_{12}$ borophene[45]. The trends of $T_c$ going with the strains are analogous to that of the values of $\lambda$, implying the variation of strain-mediated $T_c$ roots in the change of the strength of EPC. The great enhancement of $\lambda$ and $T_c$ at 5.8% can be ascribed to the Kohn anomaly induced by the huge softened phonon modes near the X point [see Fig. 6c]. Based on our results and discussion, we have partially confirmed that there is a higher chance of finding superior superconductors in the light element materials, such as in the boron hydrides discussed here. The effect of strain along the **a** direction on the superconducting properties of $\beta_{12}$-$B_5H_3$ can be found in Supplementary Note 4, which indicates the $T_c$ decreases rapidly during the tensile strain along the **a** direction increases. This denotes that despite the tensile strain along the **a** direction can trigger the Dirac semimetal phase transition but will deteriorate the superconductivity.

## Discussion

Although Dirac fermions are not uncommon in borophanes, for instance, Dirac points in $\delta_6$-borophanes[32,34] and ladder polyborane[35], Dirac nodal loops in (5-7)-$\alpha$-borophane and (5-6-7)-$\gamma$-borophane[36]; but, $\beta_{12}$-$B_5H_3$ is the first case with ideal triple Dirac fermions. On the other hand, the vanishing electron density of states in these Dirac semimetallic borophanes will deteriorate the superconducting properties, thus none of the borophanes with Dirac fermions has been reported as having superconducting features. The experimentally realized h-$B_2H_2$ is a metal, but its $T_c$ is only 11 mK[39]. It seems that it is challenging to possess Dirac fermions and superconductivity together in one borophane. In our work, the high $T_c$ (32.4 K) and clean triple Dirac fermions in $\beta_{12}$-$B_5H_3$ have proven this unreachable goal can be reached. The last but not least, $\beta_{12}$-$B_5H_3$ is not just a fantasied structure, its mother ($\beta_{12}$) and brother ($\beta_{12}$-$B_5H_2$) have been synthesized, we have sufficient confidence that it can be realized in more consideration of the good stability it otherwise has.

A few 2D borides have been claimed to harbor Dirac elements and superconductivity concurrently, such as the triple Dirac cones in 2D $AlB_6$ with a $T_c$ of 4.7 K[57], the Dirac nodal lines in bilayer $TiB_4$ with a $T_c$ of 0.82 K[58] and the Dirac nodal loop in h-$B_2O$[59] with a $T_c$ of 10.3 K[60]. But, it should be pointed out that either the Dirac states in these borides are intermixed with many other trivial electronic states near the Fermi level, not like the clean Dirac states in the $\beta_{12}$-$B_5H_3$, or their superconducting critical temperatures are much lower than the value in $\beta_{12}$-B$_5$H$_3$. The ideal Dirac properties and high $T_c$ endow the $\beta_{12}$-B$_5$H$_3$ a unique position among the 2D boron compounds.

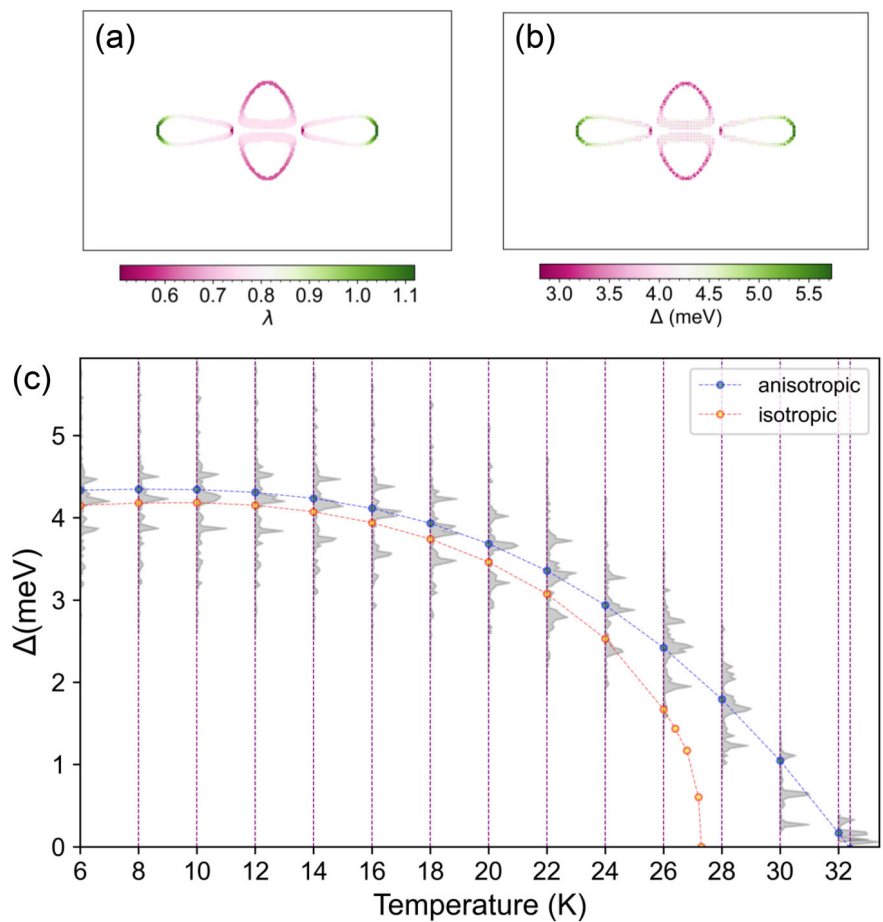

**Fig. 5 The anisotropic superconducting properties of $\beta_{12}$-B$_5$H$_3$. a** Momentum-dependent strength of electron-phonon coupling $\lambda_k$ across the full Brillouin Zone and (**b**) momentum-dependent superconducting band gap $\Delta_k$ at 10 K projected onto the Fermi surface. **c** Variation of the superconducting gap $\Delta_k$ with temperature, calculated by solving the Migdal-Eliashberg equations in the isotropic approximation (yellow dots and dashed line interpolation) and with the fully anisotropic solution, where the gray shadowed regions indicate the magnitude distribution of the $\Delta_k$ and the blue dots connected with dashed line represents the average value of the entire anisotropic $\Delta_k$.

In addition, we would like to mention that there are some other physical attributes worth being exploited in $\beta_{12}$-B$_5$H$_3$. For instance, hydrogenated borophenes have been reported to possess high thermal conductivities[61,62], which would be also expected in the $\beta_{12}$-B$_5$H$_3$; furthermore, as confirmed that the $\beta_{12}$ has a strong potential to exhibit plasmons in the visible and near-infrared regime range of frequencies with no need of doping[63], how does it is affected in the $\beta_{12}$-B$_5$H$_3$ is also an interesting question.

## conclusion
In conclusion, a hydrogenated borophene named $\beta_{12}$-B$_5$H$_3$ has been proposed that exhibits a bunch of properties. We determine its good stability. Its good elastic properties demonstrate that the material can be readily tailored by external strain. Its electronic band structure involves three clean Dirac points (two type-I Dirac points and one type-II Dirac point) near the Fermi level. A quantum phase transition in $\beta_{12}$-B$_5$H$_3$ from a Dirac semimetal with triple Dirac points to a Dirac semimetal with a single Dirac point that can be tunable by external strain along the **a** direction. Another aspect is that $\beta_{12}$-B$_5$H$_3$ is also a 2D phonon-mediated superconductor with a high $T_c$ of 32.4 K by solving the fully anisotropic ME equations. We also reveal that the $T_c$ may be slightly suppressed under the medium tensile strain along the **b** direction but eventually be enhanced to 42 K at 5.8% tensile strain along the **b** direction. The proposed 2D $\beta_{12}$-B$_5$H$_3$ thus offers a platform not only for the investigation of quantum phase transition in Dirac semimetal but also for the study of the superconductivity and the potential rich physics brought about by their interplay.

## Methods
**Structural and electronic properties.** First-principles calculations were carried out within the Vienna ab initio Simulation Package (VASP)[64,65] based on density functional theory (DFT). The exchange-correlation functional of Perdew-Burke-Ernzerhof (PBE)[66] along with the projector-augmented wave (PAW)[67] pseudopotentials were employed for the self-consistent total energy calculations and geometry optimization. The kinetic energy cutoffs were chosen to be 500 eV. Atomic positions were relaxed until the energy difference was smaller than $10^{-5}$ eV and the maximum Hellmann-Feynman forces imposed on any atoms were below $10^{-2}$ eV/Å. The Brillouin Zone (BZ) was sampled with a $12 \times 8 \times 1$ Monkhorst-Pack k-point mesh. The vacuum thick was set to 25 Å.

**Phononic and superconducting properties.** The phonon and superconducting properties were calculated in the Quantum-ESPRESSO (QE) package[51]. The PBE exchange-correlation functional and PAW pseudopotential with a 60 Ry cutoff energy were adopted to model the electron-ion interactions. The structural and electronic properties were recalculated by QE, and consistent results with those calculated by VASP were obtained. The vibrational properties and phonon perturbation potentials were calculated on a $12 \times 8 \times 1$ mesh of $q$-points within the framework of density-functional perturbation theory (DFPT)[50], combing with Methfessel-Paxton smearing scheme of width 0.02 Ry and $24 \times 16 \times 1 k$-point mesh. Once the phonon perturbation potentials were obtained in QE, then we shift our calculations to the Electron-phonon Wannier (EPW) 5.4 code[68] to solve the ME equation both in the isotropic and anisotropic approximations to obtain the superconducting gap and its temperature evolution. Fine electron ($240 \times 160 \times 1$) and phonon ($120 \times 80 \times 1$) grids were used to interpolate the EPC constant through the maximally localized Wannier functions[69] as implemented in the EPW code, where the results of electronic band structures calculated with QE and EPW can be found in Supplementary Fig. 6. In all cases, a 0.8 eV cutoff for the Matsubara frequency was chosen; the Dirac delta functions for electrons and phonons were smeared out with the widths of 15 meV and 0.3 meV, respectively; a typical value of 0.1 was used for the screened Coulomb parameter $\mu^*$.

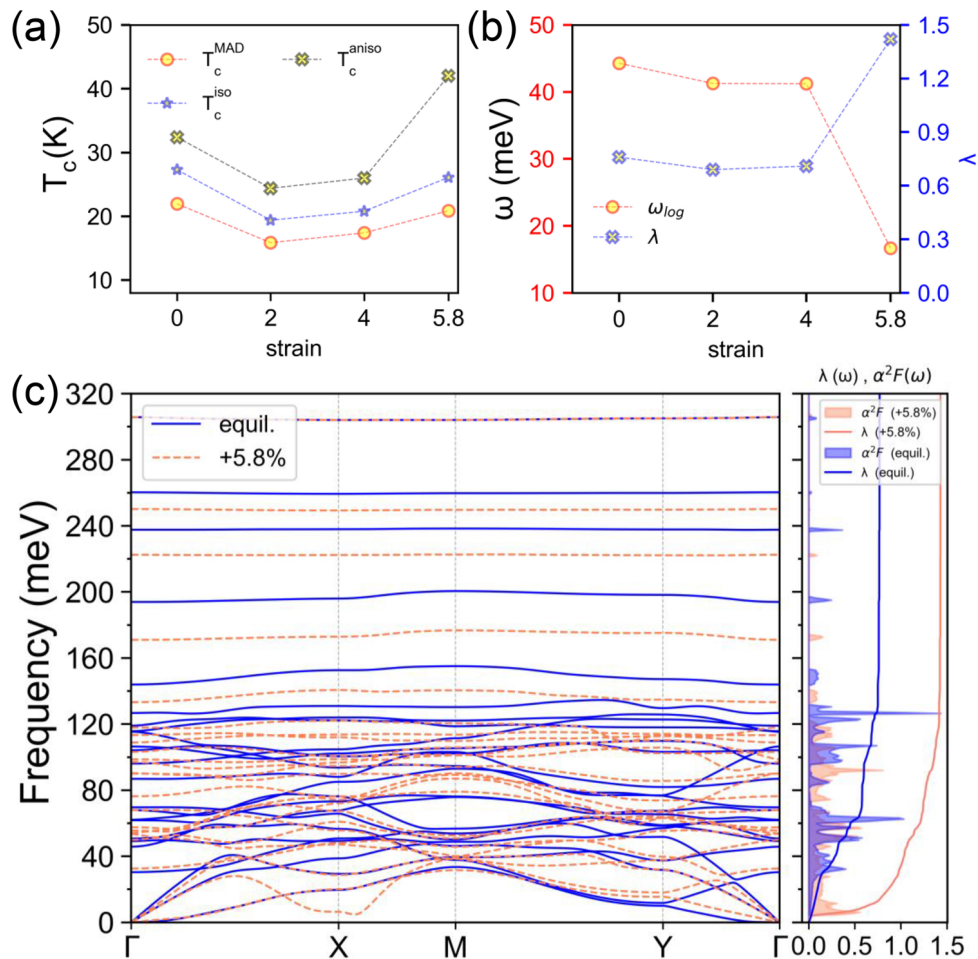

**Fig. 6 The superconducting properties of $\beta_{12}$-B$_5$H$_3$ under tensile straion along b direction. a** Evolution of superconducting transition temperature ($T_c$) as a function of tensile strain along the **b** direction at three different approximations [e.g. the McMillian-Allen-Dynes(MAD), isotropic Migdal-Eliashberg (iso) and anisotropic Migdal-Eliashberg (aniso)]. **b** Evolution of electron-phonon coupling (EPC) $\lambda$ and the logarithmic average frequency $\omega_{\log}$. **c** Phonon band structure, isotropic Eliashberg spectral function $\alpha^2 F$ and EPC $\lambda$ under 5.8% tensile strain along the **b** direction. Data in the equilibrium (equil.) case are shown by the blue shadow and line for comparison.

### Data availability
The structural details of $\beta_{12}$-B$_5$H$_3$ is provided in the Supplementary Note 5. The other data that support the findings of this study are available from the corresponding author (Chengyong Zhong) upon reasonable request.

### Code availability
The Vienna Ab Initio Simulation Package is a proprietary software available for purchase at https://www.vasp.at/. Quantum ESPRESSO is open-source, the latest stable version can be downloaded at https://www.quantum-espresso.org/. Electron-phonon Wannier code is distributed as part of the Quantum ESPRESSO materials simulation suite.

## Acknowledgements

C.Z. acknowledges the financial support from the Scientific and Technology Research Program of Chongqing Municipal Education Commission (KJQN202300515) and the Foundation of Chongqing Normal University (23XLB001). We also wish to acknowledge the financial support from the National Natural Science Foundation of China (No. 11804039 and 52071043).

### Author contributions

C.Z. conceived the project, performed all of the calculations, wrote and revised the manuscript. X.L. helped to analyze some of the data. P.Y. provided the funding. All authors modified and approved the manuscript.

### Competing interests

The authors declare no competing interests.

### Additional information

**Supplementary information** The online version contains supplementary material available at https://doi.org/10.1038/s42005-024-01523-x.

**Correspondence** and requests for materials should be addressed to Chengyong Zhong or Peng Yu.

**Peer review information** *Communications Physics* thanks the anonymous reviewers for their contribution to the peer review of this work.

**Reprints and permission information** is available at http://www.nature.com/reprints

**Publisher's note** Springer Nature remains neutral with regard to jurisdictional claims in published maps and institutional affiliations.

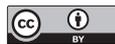